# Topological chiral edge states in deep-subwavelength valley photonic metamaterials

*Rui Xi[1,2,#], Qiaolu Chen[1,2,#,]\*, Qinghui Yan[1,2], Li Zhang[1,2], Fujia Chen[1,2], Ying Li[1,2], Hongsheng Chen[1,2,]\*, and Yihao Yang[1,2,]\**

R. Xi, Q. Chen, Q. Yan, L. Zhang, F. Chen, Y. Li, H. Chen, and Y. Yang
Interdisciplinary Center for Quantum Information, State Key Laboratory of Modern Optical Instrumentation, ZJU-Hangzhou Global Scientific and Technological Innovation Center, Zhejiang University, Hangzhou 310027, China.
Emails: qiaoluchen@zju.edu.cn (Q. C.); hansomchen@zju.edu.cn (H. C.); yangyihao@zju.edu.cn (Y. Y.).

R. Xi, Q. Chen, Q. Yan, L. Zhang, F. Chen, Y. Li, H. Chen, and Y. Yang
International Joint Innovation Center, Key Lab. of Advanced Micro/Nano Electronic Devices & Smart Systems of Zhejiang, The Electromagnetics Academy at Zhejiang University, Zhejiang University, Haining 314400, China.
Emails: qiaoluchen@zju.edu.cn (Q. C.); hansomchen@zju.edu.cn (H. C.); yangyihao@zju.edu.cn (Y. Y.).

Abstract: Topological valley photonics has emerged as a new frontier in photonics with many promising applications. Previous valley boundary transport relies on kink states at internal boundaries between two topologically distinct domains. However, recent studies have revealed a novel class of topological chiral edge states (CESs) at external boundaries of valley materials, which have remained elusive in photonics. Here, we propose and experimentally demonstrate the topological CESs in valley photonic metamaterials (VPMMs) by accurately tuning on-site edge potentials. Moreover, the VPMMs work at deep-subwavelength scales. Thus, the supported CESs are highly confined and self-guiding without relying on a cladding layer to prevent leakage radiation. Via direct near-field measurements, we observe the bulk bandgap, the edge dispersions, and the robust edge transport passing through sharp corners, which are hallmarks of the CESs. Our work paves a way to explore novel topological edge states in valley photonics and sheds light on robust and miniaturized photonic devices.

## 1. Introduction

Originated from condensed-matter physics, the valley has provided a fundamentally novel degree of freedom (DOF) beyond amplitude, phase, and polarization, to manipulate light in photonics.[1-7] The valley refers to the local minimum of the conduction bands or local maximum of the valence bands that usually locates at the corners of the Brillouin zone, as widely exemplified by the graphene-like structures (i.e., analogies of the graphene with staggered sublattice potentials),[8] such as photonic crystals, coupled optical waveguide arrays, and others.[9-25] A remarkable property of valley photonic materials is that they support robust valley kink states (also known as zero modes) at domain walls between two topologically distinct domains, which can bypass sharp corners and impurities with negligible inter-valley scattering. Due to their intriguing properties, the valley photonic materials have shown great potentials in practical applications, ranging from on-chip communications,[21,22,24] quantum chiral optics,[26-27] to robust lasers.[28-31]

Recent advances have suggested a new type of edge state existing at the external edges of the valley materials by tuning the on-site edge potentials.[32-33] These edge states share the same topological origin as the previous valley kink states and exhibit chiral nature related to the valley-dependent bulk topological charges. Therefore, such chiral edge states (CESs) show valley-momentum locking properties and topologically enabled robustness and could find many applications as the valley kink states. In contrast to the valley kink states that rely on domain-wall configurations or internal boundaries, the CESs appear truly at edges or external boundaries of the valley materials, enabling more compact photonic devices. The valley-polarized CESs are very similar to the well-known spin-polarized edge states in quantum spin Hall insulators,[34-35] with the valleys playing the role of spins. However, such topological CESs have remained elusive in photonics.

When implementing the topological CESs in photonics, one inevitably encounters a big challenge: for photons, the vacuum or air behaves as a "conductor", giving rise to unavoidable radiation leakage to free space, especially when the photons' momenta at edges and in free space are comparable.[36] Note that some physical settings do not have the above problem; for example, the vacuum behaves as an insulator for electrons or phonons in solids. Some metals or artificial photonic bandgap materials are insulating for photons; thus, they can serve as a cladding layer to confine photons at the edges. However, the involvement of the cladding layer undermines the benefits of the topological CESs. Moreover, the cladding layer may alter the on-site edge potentials, imposing new challenges to the implementation.

Given the above demands, we propose and experimentally realize the topological CESs in valley photonic metamaterials (VPMMs) by accurately tuning the on-site edge potentials. Moreover, the designed VPMMs work at deep subwavelength scales--the unit cell size is around $\lambda_0/17$, with $\lambda_0$ being the operational free-space wavelength. Thus, the CESs are extremely confined and self-guiding, protected from radiation leakage. Such significantly squeezed CESs are also beneficial to miniaturization and integration of the topological photonic devices, which have been highly pursued but never been realized so far in valley photonics.[9,11,13-16,20-25] Via direct near-field measurements, we experimentally visualize the bulk dispersions of the VPMM and the dispersions of the CESs. We further demonstrate the robust CESs passing through 120° sharp corners and smooth transition between two distinct CESs.

## 2. Results

We start with a metamaterial in a honeycomb lattice (see **Figure 1**a and Figure1b), where each unit cell includes two triangular coiling wires with winding numbers $n_2 = n_0 - \delta n_b$ and $n_3 = n_0 + \delta n_b$, respectively (see the right panel of Figure 1b). The metamaterial has a period of $a = 13.16$ mm, and the metallic coil has a gap of $g = 0.4$ mm and a width of $w = 0.4$ mm, patterned on a substrate with a thickness of $t = 2$ mm and a relative permittivity of $\varepsilon = 2.1$.

When $\delta n_b = 0$, our metamaterial possesses the inversion symmetry and time-reversal symmetry, hence generating a pair of Dirac points around K and K' valleys at 1.36 GHz. Note that the Dirac points are slightly deviated from the K and K' points due to the nearly-$C_6$ symmetry (see more details in Supporting Information). For simplicity, we define the momenta of the Dirac degeneracy as $K_1$ and $K_1'$ points. Upon breaking the inversion symmetry (for example, $\delta n_b = 0.17$), the Dirac degeneracy is lifted, resulting in a bulk bandgap (1.31 GHz $< f <$ 1.46 GHz), as depicted by the magenta curves in Figure 1c. The low-energy Hamiltonian near $K_1$ valley can be written as $\delta H_{K_1}(\delta k) = v_D(\delta k_x \sigma_x + \delta k_y \sigma_y) + m v_D^2 \sigma_z$, where $\delta \boldsymbol{k} = \boldsymbol{k} - \boldsymbol{k_1}$ is the displacement of wave vector $\boldsymbol{k}$ to $K_1$ valley in momentum space, $v_D$ is the group velocity, $\sigma_i$ ($i = x, y, z$) are elements in the Pauli matrices, and $m$ is the effective mass term induced by the broken inversion symmetry. We also numerically derive the Berry curvatures around $K_1$ and $K_1'$ points based on the first-principles calculation, which lead to non-trivial valley Chern numbers $\pm 1/2$ (see the Supporting Information).[13,15,22] It has been known that the valley kink states exist at the internal boundaries between two domains with opposite non-trivial valley Chern numbers. In the following, we focus on the realization of valley-locked CESs appearing at edges or external boundaries of our VPMM [32-33,37].

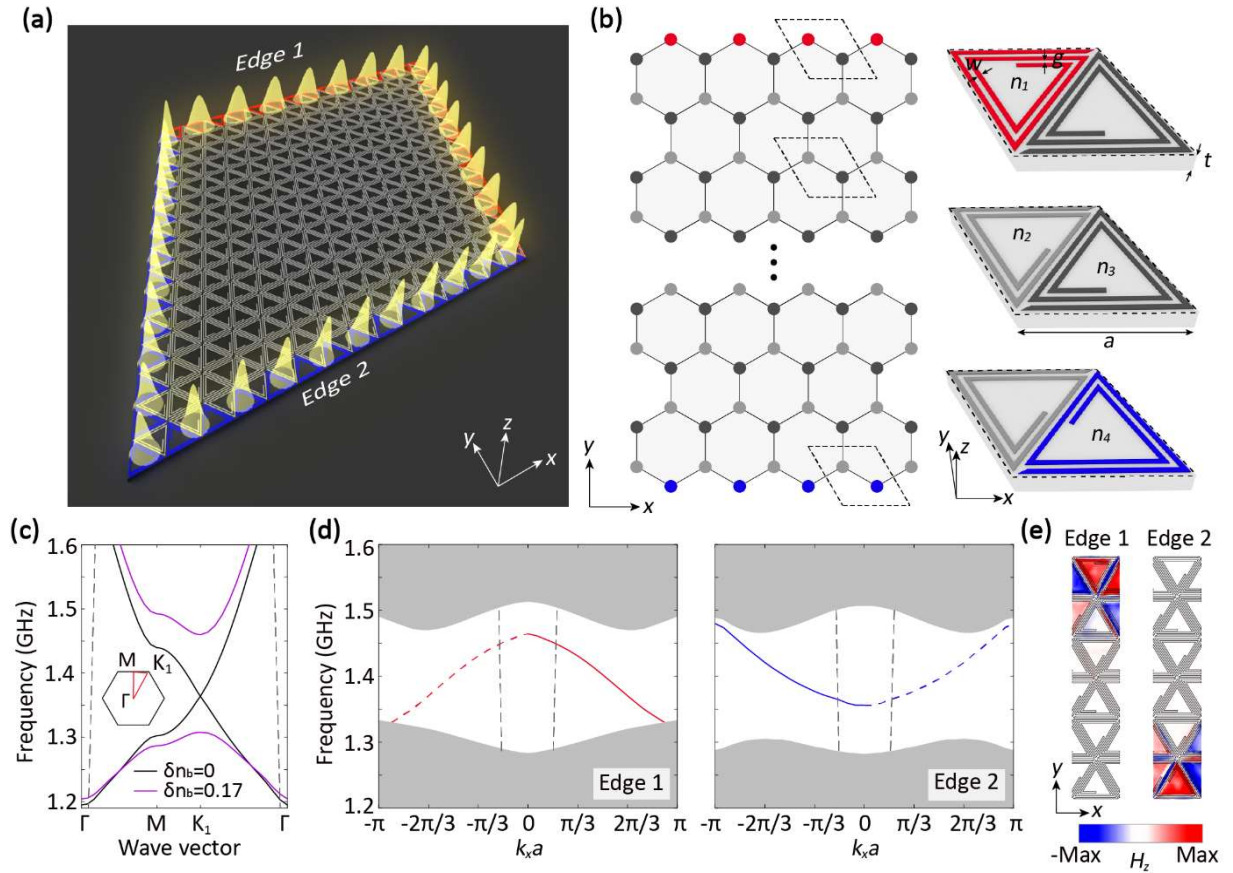

**Figure 1.** Topological CESs at external edges of the VPMM. (a) Schematic view of the proposed VPMM with Edge 1 (red) and Edge 2 (blue). (b) Schematic view of the proposed honeycomb lattice with broken inversion symmetry. Black dashed rhombuses display a bulk unit cell containing two sublattices denoted by grey and black dots, and an outmost unit cell at Edge 1 (Edge 2) depicted by red and black (grey and blue) dots. Right panels: the bulk unit cell with winding numbers $n_2 = n_0 - \delta n_b$ and $n_3 = n_0 + \delta n_b$; the outmost coil at Edge 1 with winding numbers $n_1 = n_0 - \delta n_b + \delta n_{e1}$ and $n_3 = n_0 + \delta n_b$; the outmost coil at Edge 2 with winding numbers $n_4 = n_0 + \delta n_b - \delta n_{e2}$ and $n_2 = n_0 - \delta n_b$. Here, $a$ = 13.16 mm, $w = g$ = 0.4 mm, $t$ = 2 mm, $n_0$ = 2, $\delta n_b$ = 0.17, $\delta n_{e1}$ = 0.37, and $\delta n_{e2}$ = 0.33. (c) Band diagram of the bulk region of the metamaterial with inversion symmetry (black curves, $\delta n_b = 0$) and broken inversion symmetry (magenta curves, $\delta n_b$ = 0.17), displaying a Dirac point at $K_1$ point, and a bulk bandgap from 1.31 to 1.46 GHz. Inset is the first Brillouin zone, labelling $\Gamma$, $K_1$, and $M$ points. Gray dashed curves represent the light line. (d) Band diagrams of Edge 1 and Edge 2, respectively. The shaded regions denote the projections of bulk bands. Dashed (solid) curves represent the CESs with positive (negative) group velocities. Gray dashed curves display the light line. (e) Simulated field maps of $H_z$ component, showing the CESs at Edge 1 and Edge 2, respectively.

According to Ref. [32, 33], to realize the topological CESs at the external boundaries of VPMM, the boundary matrix should satisfy $M = \tau_z \sigma_y$, where $\tau_i (i = x, y, z)$ are the Pauli matrices. Here, the boundary matrix $M$ phenomenologically describes the boundary conditions at the edges (see the Supporting Information). The required boundary matrix can be achieved by tuning the on-site edge potentials of the zigzag edges, whose effective boundary matrix is $M_{zigzag} = \tau_z (\sigma_y \sin\theta_V + \sigma_z \cos\theta_V)$ with $\theta_V$ governed by the on-site edge potentials.[32-33] When $\theta_V = \pi/2$, $M_{zigzag} = M$. Following the above theoretical analysis, we alter the on-site edge potentials by changing the winding numbers of the outmost coils at zigzag-terminated edges, namely Edge 1 and Edge 2, as illustrated schematically in Figure 1b. Via numerical calculations, we indeed observe that the dispersions of Edge 1 and Edge 2 vary as the winding number of the outmost coil changes (see the Supporting Information). The optimal winding numbers of the outmost coils at Edge 1 and Edge 2 are $n_1$ = 2.20 and $n_4$ = 1.84, respectively (see the right panel of Figure 1b).

Figure 1d displays the edge dispersions at Edge 1 and Edge 2, where the shaded regions denote the projections of the bulk bands. One can see that the CESs (i.e., the nearly-linear dispersion curves at $K_1/K_1'$ valley) occur at both edges. Interestingly, the group velocities of the CESs at $K_1$ and $K_1$' valleys are exactly opposite, which is a feature of valley-momentum locking property.[9,11,13-16,20-25] The simulated field maps of $H_z$ component are plotted in Figure 1e, showing that the CESs are highly confined at Edge 1 and Edge 2. Besides, it has been revealed that the coiling wires can support deeply subwavelength modes, where the surface currents flowing along the spiral coils produce strong magnetic dipole moments vertical to the coils [38-40]. Therefore, our coil-based VPMM operates at deep subwavelength scales, whose size of the unit cell is only about $\lambda_0/17$, much smaller than those of the previously demonstrated valley materials.[9-25] Such a deep-subwavelength nature is beneficial to the miniaturization of various topological photonic devices.

In the following, we perform the experiments to characterize intriguing properties of the topological CESs. The experimental samples are fabricated using the standard printed circuit board (PCB) etching techniques (see **Figure 2**a, **Figure 3**a and **Figure 4**a). The coiling copper wires are patterned on a dielectric substrate with a relative permittivity of $\varepsilon = 2.1$ and a loss tangent of 0.003. In the experimental setup, a port of a vector network analyzer (VNA) is directly connected to an outmost coil of the experimental sample to excite the modes efficiently, and another port is connected to a triangular coil with a magnetic resonance around 1.4 GHz, serving as a detector. By fixing the coil-like detector on a robotic arm of a movement platform, the field distributions, including amplitudes and phases over the experimental sample, can be mapped out (see the details in the Supporting Information).

We first experimentally characterize the bulk and edge dispersions of the sample shown in Figure 2a. To selectively excite the CESs and the bulk states, we place the sources at different locations marked by red, blue and grey dots in Figure 2a, respectively. Figure 2b plots the amplitude ratios of the scattering parameter $S_{21}$ between the output and input ports, which are proportional to $H_z$ field intensity. We observe an approximately 30-dB drop in the transmissions of the bulk states from 1.31 to 1.47 GHz (black curve) that corresponds to a bulk bandgap (shaded area). Along Edge 1 and Edge 2, however, the measured transmissions (red and blue curves) keep high throughout this frequency region, revealing the existence of the topological CESs. Besides, to probe the bulk band diagram, we scan the $H_z$ fields on the *xy* plane above the sample and then apply Fourier transform to the real-space complex field distributions of the bulk region. As shown in Figure 2c, the measured projected band diagram is highly consistent with the numerical results (green curves).

Next, we directly visualize the CESs by mapping the $H_z$ fields. Figure 2d displays the measured $H_z$ field distributions at Edge 1 and Edge 2 at 1.42 GHz, in which the topological CESs are strongly confined at both edges. For reference, we plot the simulated $H_z$ field maps in Figure 2e, which match well with the experimental results. We also perform the spatial Fourier transform to the measured $H_z$ fields around two edges, as illustrated in Figure 2f. It is evident that the measured dispersions have a good agreement with the simulated counterparts (green curves in Figure 2f), corroborating the existence of the CESs.

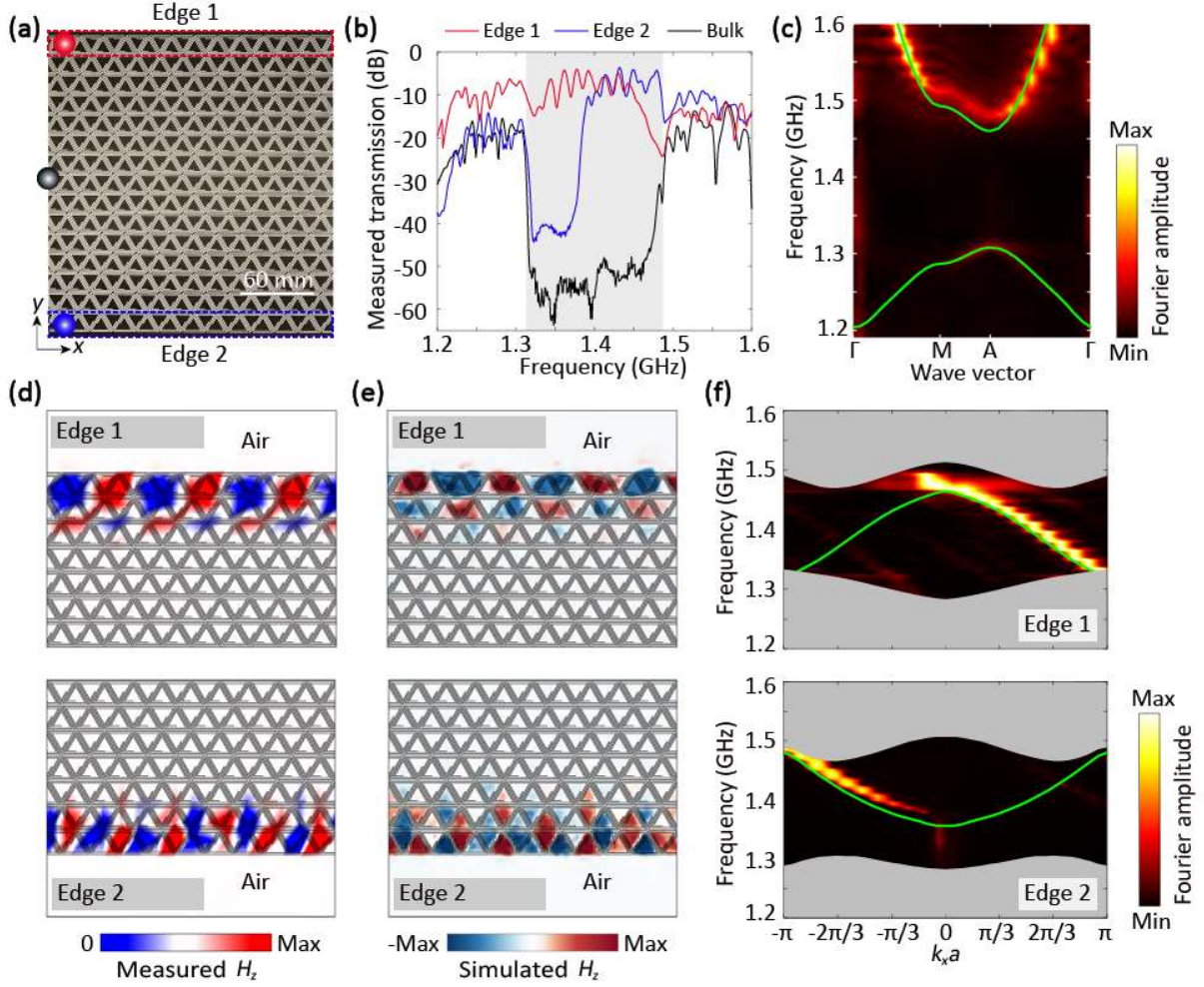

**Figure 2.** Observation of the topological CESs. (a) Top-view photograph of the VPMM with two zigzag edges, i.e., Edge 1 marked by a red dashed rectangle and Edge 2 denoted by a blue dashed rectangle. Red, blue, and grey dots represent excitations for CESs and bulk states, respectively. (b) Measured transmissions of the bulk states (black curve) and the CESs (red and blue curves). The shaded area represents the bulk bandgap. (c) Measured bulk dispersions of the VPMM, obtained by applying Fourier transform to the real-space field distributions of the bulk region. The color scale measures the magnetic energy density $|H|$. Green curves correspond to simulated dispersions. (d), (e) Measured (d) and simulated (e) $H_z$ field maps near Edge 1 (top panels) and Edge 2 (bottom panels) at 1.42 GHz. The color scale measures the amplitude of $H_z$ field. (f) Measured edge dispersions by applying Fourier transform to the real-space field distributions around Edge 1 (top panel) and Edge 2 (bottom panel). Green curves correspond to the simulated dispersions. The color scale measures the magnetic energy density $|H|$.

Since the CESs share the exactly same topological origin as the valley kink states,[33] both of them are robust against disorders, such as sharp corners. To verify this point, we fabricate a sample shown in Figure 3a, which consists of three edges (Edge 1) and three 120° sharp turns. Figure 3b displays the measured transmissions of the CESs along the sharp corners, which are comparable with those measured along the previous Edge 1 within the bandgap frequency region, unveiling a key feature of the topological CESs. Note that due to the metallic loss of the experimental sample, it is hard to recognize the regularly spaced peaks of topological cavity modes in the spectrum intensities.[25,31]

We further experimentally map out the $H_z$ field distributions over the triangle-shaped sample. As shown in Figure 3c, the energy uniformly distributes at the external boundaries, confirming the robust propagation of the CESs against the sharp turns. This experimental result is almost identical to the numerical one (Figure 3d). Thus, the topological cavity based on the robust CESs could be a novel compact platform for the topological lasing.[28-31]

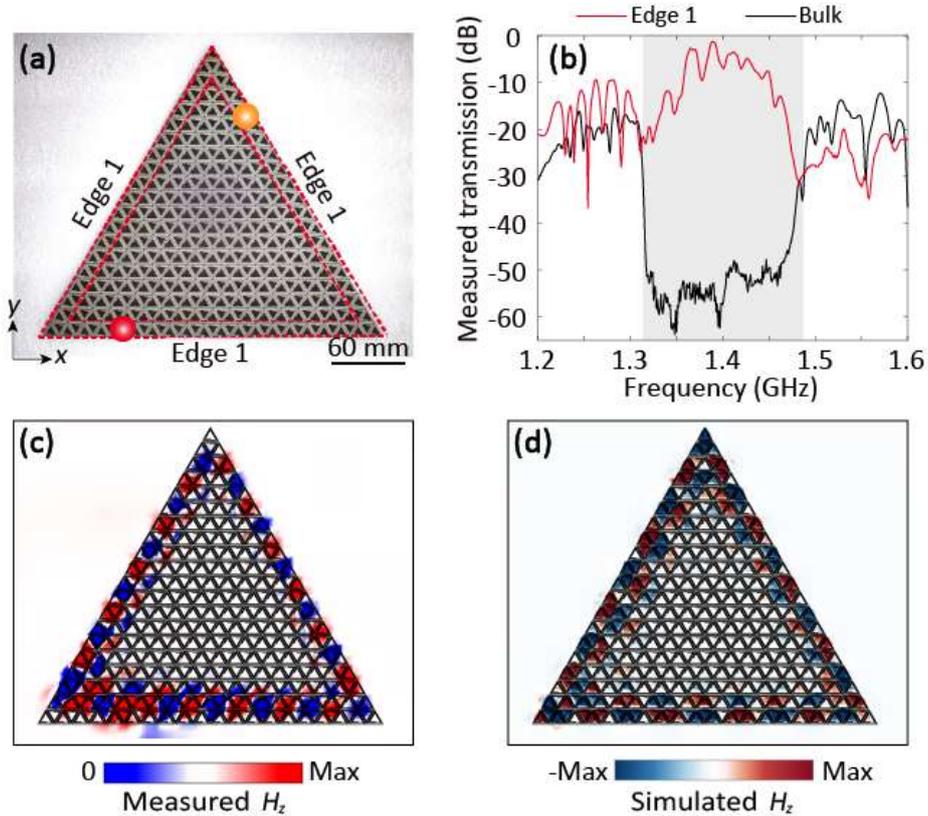

**Figure 3.** Robust topological CESs passing through three 120° sharp corners. (a) Top-view photograph of the triangle-shaped sample containing three Edge-1 boundaries and three 120° sharp turns. The Red (orange) dot represents the excitation (detector). (b) Measured transmissions of the CESs (red curve) along three 120° sharp turns. Black curve: the bulk transmissions. Shaded area: the bulk bandgap. (c), (d) Measured (c) and simulated (d) $H_z$ field maps in the triangle-shaped sample at 1.41 GHz. The color scale measures the amplitude of $H_z$ field.

We also experimentally demonstrate the smooth transition between two CESs at Edge 1 and Edge 2. As illustrated in Figure 4a, Edge 1 and Edge 2 intersect at a 60° corner. A source is placed on the left side of Edge 2 (blue dot) to launch the CESs. From the $S_{21}$ parameter in Figure 4b, one can see that the measured transmissions along the twisted boundaries remain high throughout the frequency region where two CESs at Edge 1 and Edge 2 exist simultaneously (red dashed area). We further map out the field distributions at 1.44 GHz. The experimental and simulated results are shown in Figure 4c and Figure 4d, respectively, indicating that the topological CESs are tightly confined at the external boundaries and transport robustly and smoothly around a 60° corner from Edge 2 to Edge 1, with negligible reflection.

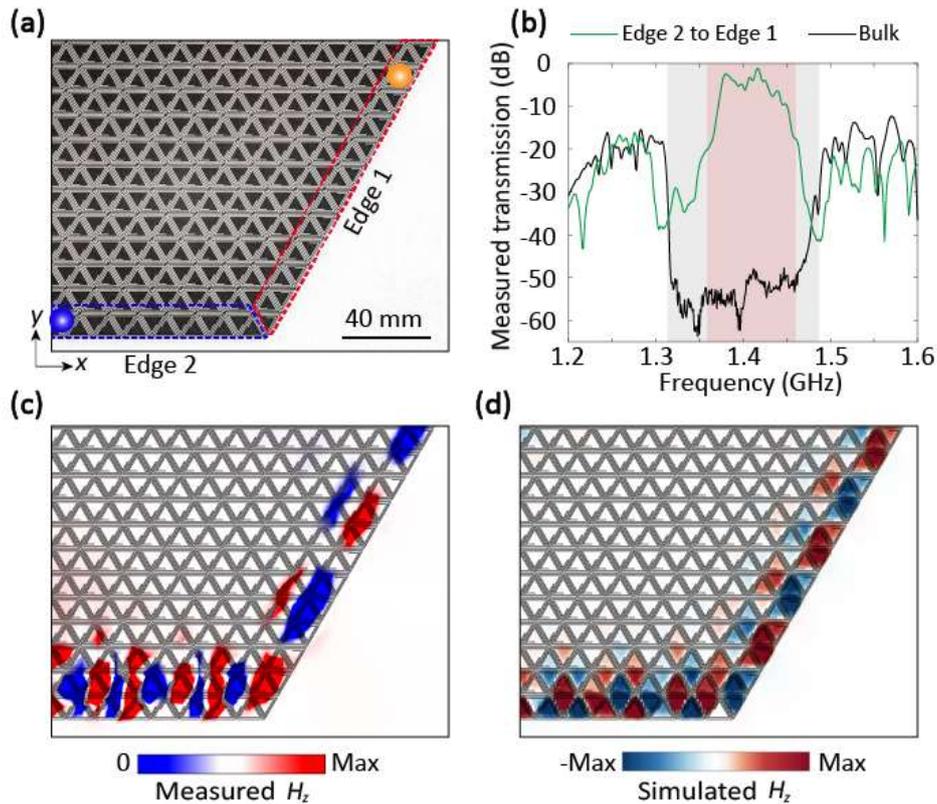

**Figure 4.** Smooth transition between two topological CESs at Edge 1 and Edge 2. (a) Top-view photograph of the fabricated sample containing Edge 1 and Edge 2. The blue (orange) dot represents the excitation at Edge 2 (detector at Edge 1). (b) Measured transmissions of the CESs (green curve) propagating from Edge 2 to Edge 1. Black curve: the bulk transmissions. Gray shaded area: the bulk bandgap. Red shaded area: the frequency region where the CESs at Edge 1 and Edge 2 exist simultaneously. (c), (d) Measured (c) and simulated (d) $H_z$ field maps around the transition region at 1.44 GHz. The color scale measures the amplitude of $H_z$ field.

## 3. Discussions

We have thus theoretically proposed and experimentally realized the robust CESs at the external boundaries of VPMMs by precisely tuning the on-site edge potentials. Via direct near-field measurements, we demonstrate the bulk and edge dispersions, the robust transport through 120° sharp turns, and the smooth transition between two CESs. Such self-guiding CESs do not rely on internal boundaries or cladding layers to prevent radiation leakage, which can be used to design photonic devices with smaller footprints than those based on the conventional valley kink states. Besides, we expect that the realization of photonic CESs by tuning the on-site edge potentials could stimulate theoretical and experimental studies on novel photonic topological edge states by tailoring the unit cells globally or locally. Finally, the highly-confined CESs can find applications in miniaturized topological photonic devices, such as robust delay lines,[41] robust on-chip communications, and topological lasers.[28-31]

## 4. Methods

**Numerical Simulations:**

We numerically simulate the bulk diagrams of the VPMMs in a finite-element method solver (COMSOL Multiphysics). The simulations of the topological CESs are performed in the eigenvalue module of CST Microwave Studio. For bulk (edge) dispersions calculations, periodic boundary conditions are applied on $x$ and $y$ axes of the unit cell (supercell) to form an infinite hexagonal (rectangular) lattice. The field patterns are performed in the time domain of CST Microwave Studio. We also consider the copper and the substrate as a perfect electric conductor (PEC) and a lossless medium in the simulations, respectively.

**Experimental samples:**

The experimental samples are fabricated with the PCB technique by etching 0.035-mm-thick copper cladding on 2-mm-thick dielectric substrates. One rectangular sample consists of Edge 1 and Edge 2 to support the bulk and edge transport. Besides, a triangular sample with Edge-1 boundaries and a rhombus sample with a 60-degree junction between Edge 1 and Edge 2 is fabricated.

Feeding microstrips are fabricated along the external boundaries of the samples. A triangular coil-like detector is designed and fabricated with a magnetic resonance around 1.4 GHz to probe the excited modes (Figure S1, Supporting Information).

**Measurements:**

In the measurement, a VNA (R&S ZVL13) is utilized to measure the amplitude and phase of the magnetic field. Two ports of the VNA are connected with a source and a detector, respectively. The coil-like detector is fixed at a robotic arm of a 3D movement platform (Linbou NFS03). The field distributions over the fabricated samples can be mapped out by probing both the amplitude and phase of the magnetic field point by point (Figure S1, Supporting Information).

**Supporting Information**
Supporting Information is available from the Wiley Online Library.

**Acknowledgements**
Rui Xi and Qiaolu Chen contributed equally to this work. The work was sponsored by the National Natural Science Foundation of China (NNSFC) under Grants No. 61625502, No. 62175215, No.11961141010, and No. 61975176, the Top-Notch Young Talents Program of China, the Fundamental Research Funds for the Central Universities.

**Conflict of Interest**
The author declares no conflict of interest.

**Data Availability Statement**
All data are available from the corresponding author upon reasonable request.